\documentclass[prl,twocolumn,aps,superscriptaddress,showpacs,floatfix]{revtex4}
\usepackage{amssymb}
\usepackage{amsmath}
\usepackage{graphicx}
\usepackage[normalem]{ulem}
\usepackage[dvips]{color}

\begin{document}

\title{Nuclear symmetry energy and its density
slope at normal density extracted from global nucleon optical
potentials}

\author{Chang Xu}
\affiliation{Department of Physics and Astronomy, Texas A$\&$M
University-Commerce, Commerce, Texas 75429-3011,
USA}\affiliation{Department of Physics, Nanjing University,
Nanjing 210008, China}
\author{Bao-An Li\footnote{Corresponding author, Bao-An\_Li$@$Tamu-Commerce.edu}}
\affiliation{Department of Physics and Astronomy, Texas A$\&$M
University-Commerce, Commerce, Texas 75429-3011, USA}
\author{Lie-Wen Chen}
\affiliation{Department of Physics and Astronomy, Texas A$\&$M
University-Commerce, Commerce, Texas 75429-3011,
USA}\affiliation{Department of Physics, Shanghai Jiao Tong
University, Shanghai 200240, China}

\begin{abstract}
Based on the Hugenholtz-Van Hove theorem, it is shown that both
the symmetry energy E$_{sym}(\rho)$ and its density slope
$L(\rho)$ at normal density $\rho_0$ are completely determined by
the global nucleon optical potentials that can be extracted
directly from nucleon-nucleus scatterings, (p,n) charge exchange
reactions and single-particle energy levels of bound states.
Adopting a value of $m^*/m=0.7$ for the nucleon effective k-mass
in symmetric nuclear matter at $\rho_0$ and averaging all
phenomenological isovector nucleon potentials constrained by world
data available in the literature since 1969, the best estimates of
$E_{sym}(\rho_0)=31.3$ MeV and $L(\rho_0)=52.7$ MeV are
simultaneously obtained. Uncertainties involved in the estimates
are discussed.

\end{abstract}


\maketitle

Nuclear symmetry energy $E_{sym}(\rho)$, which encodes the energy
related to neutron-proton asymmetry in the equation of state of
nuclear matter, is a fundamental quantity currently under intense
investigation in both nuclear physics and astrophysics
\cite{dan,JML04,AWS05,li2,li1,bro,bar,Sum94,Hor01,Sto03,tsa2,Leh09,ZGX08,Wen09,Joe10,Lee10}.
Both the magnitude and density dependence of $E_{sym}(\rho)$ are
critical for understanding not only the structure of rare isotopes
and the reaction mechanism of heavy-ion collisions, but also many
interesting issues in astrophysics, such as the structure and
composition of neutron stars. Despite much effort made both
experimentally and theoretically, our current knowledge about the
$E_{sym}(\rho)$ is still rather poor even around the saturation
density of nuclear matter $\rho_0$. More specifically, near
$\rho_0$ the symmetry energy can be characterized by using the
value of $E_{sym}(\rho_0)$ and the slope parameter $L(\rho_0) = 3
\rho_0 \frac{\partial E_{sym}(\rho)}{\partial
\rho}|_{\rho=\rho_0}$, \textit{i.e.},
\begin{eqnarray}
E_{sym}(\rho)=E_{sym}(\rho_0)+\frac{L(\rho_0)}{3}(\frac{\rho-\rho_0}{\rho_0})
+O((\frac{\rho-\rho_0}{\rho_0})^2).
\end{eqnarray}
The $E_{sym}(\rho_0)$ is known to be around $28-34$ MeV mainly
from analyzing nuclear masses within liquid-drop models
\cite{mye63,Peter95,pomo}. The exact value of $E_{sym}(\rho_0)$
extracted in this way depends largely on the size and accuracy of
nuclear mass data base as well as whether/what surface symmetry
energy is used in the analysis. The empirical value of
$E_{sym}(\rho_0)$ has been used in calibrating many-body theories
with various interactions. The density slope $L(\rho_0)$ has also
been extensively studied but remains much more uncertain. Since
its exact value is particularly important for determining several
critical quantities, such as, the size of neutron-skin in heavy
nuclei \cite{Bro00,Fur02,Die03,Che05b,Cen09,Sam09}, location of
neutron dripline \cite{Kaz10}, core-crust transition density
\cite{Hor01,JML04,AWS05,Oya07,Kub07,XCLM09,Mou10,Duc10} and
gravitational binding energy \cite{Newton09} of neutron stars,
much more effort has been devoted recently to extracting the value
of $L$ from studying various phenomena/observables in terrestrial
nuclear laboratory experiments. These include the isospin
diffusion, neutron/proton ratio of pre-equilibrium nucleon
emissions and isoscaling in heavy-ion reactions
\cite{LWC05,tsa1,tsa2,Fami,Ste05,shet}, energy shift of isobaric
analogue states \cite{Pawel09}, neutron-skin of heavy nuclei
\cite{Cen09,Chen10}, and giant dipole as well as pygmy dipole
resonances \cite{Tri08,Kli07,Car10}. Unfortunately, the values of
$L$ extracted so far from these studies scatter between about 20
to 115 MeV. Since all these phenomena/observables are in someway
at least indirectly and qualitatively related to the $L$
parameter, it is very useful to know if one can directly express
the $L$ in terms of some relevant parts of the commonly used
underlying nuclear effective interaction. In this Letter, based on
the Hugenholtz-Van Hove (HVH) theorem \cite{hug}, it is firstly
shown analytically that the $L(\rho)$ is determined completely by
the momentum-dependent global nucleon optical potential that can
be directly extracted from nucleon-nucleus and (p, n) charge
exchange reactions. By averaging all phenomenological global
isovector optical potentials constrained by the world data
available in the literature since 1969, the best estimates of
E$_{sym}(\rho_0)=31.3$ MeV and $L(\rho_0)=52.7$ MeV are obtained
simultaneously.

The general HVH theorem \cite{hug}, \textit{i.e.},
\begin{equation}
E_{F}=\frac{d\xi}{d \rho}=\frac{d (\rho E_{av})}{d \rho} = E_{av} +
\frac{P}{\rho},
\end{equation}
is a fundamental relationship among the Fermi energy $E_{F}$,
energy density $\xi=\rho E_{av}$ where $E_{av}$ is the average
energy per nucleon and the pressure of the system $P$ at zero
temperature. The HVH theorem has been strictly proven to be valid
for any interacting self-bound infinite Fermi system and is
independent of the precise nature of the interaction used. At
normal density $\rho_0$ where the pressure vanishes ($P=0$), the
general HVH theorem reduces to the special relation $E_F=E_{av}$.
According to the general HVH theorem, the Fermi energies of
neutrons and protons in nuclear matter of isospin asymmetry
$\delta=(\rho_n-\rho_p)/\rho$ at an arbitrary density $\rho$ are,
respectively,
\begin{eqnarray}
t(k_F^n)+U_n(\rho,\delta,k_F^n) = \frac{\partial \xi }{\partial
\rho_n}, \label{chemUn}
\\
t(k_F^p)+U_p(\rho,\delta,k_F^p) = \frac{\partial \xi }{\partial
\rho_p}, \label{chemUp}
\end{eqnarray}
where $t(k_F^{n/p})$ and $U_{n/p}$ are the neutron/proton kinetic
energy and single-particle potential, respectively. The Fermi
momenta of neutrons and protons are $k_F^n=k_F(1+\delta)^{1/3}$ and
$k_F^p=k_F(1-\delta)^{1/3}$, respectively, with
$k_F=(3\pi^2\rho/2)^{1/3}$. According to the well-known Lane
relationship \cite{Lan62}, the $U_{n/p}$ can be well approximated by
$U_{n/p}(\rho,\delta,k) \simeq U_0(\rho,k) \pm
U_{sym}(\rho,k)\delta$, where $U_0(\rho,k)$ and $U_{sym}(\rho,k)$
are, respectively, the isoscalar and isovector (symmetry) nucleon
potentials. While the $U_{sym}(\rho,k)$ still has significant
uncertainties the $U_0(\rho,k)$ especially at $\rho_0$ has been
relatively well determined \cite{li2,dan,li1}. By subtracting the
Eq.(\ref{chemUp}) from Eq.(\ref{chemUn}) and then making a Taylor's
expansion in $\delta$ of both the $t(k_F^{n/p})$ and
$U_{n/p}(\rho,\delta,k_F^{n/p})$, one can show that \cite{bru64,xu1}
\begin{eqnarray}
E_{sym}(\rho)= \frac{1}{6} \frac{\partial (t+U_0)}{\partial
k}|_{k_F}k_F + \frac{1}{2} U_{sym}(\rho,k_F). \label{Esymexp}
\end{eqnarray}
Moreover, by firstly adding the Eq.(\ref{chemUn}) with
Eq.(\ref{chemUp}) and then expanding in $\delta$ again both the
$t(k_F^{n/p})$ and $U_{n/p}(\rho,\delta,k_F^{n/p})$ while noticing
the Eq.(\ref{Esymexp}), we obtain for the first time that
\begin{eqnarray}
L(\rho) &=& \frac{1}{6} \frac{\partial (t+U_0)}{\partial k}|_{k_F}
\cdot k_F + \frac{1}{6} \frac{\partial^2 (t+U_0)}{\partial
k^2}|_{k_F} \cdot
 k_F^2  \nonumber \\
 & + & \frac{3}{2} U_{sym}(\rho,k_F)
  + \frac{\partial U_{sym}}{\partial k}|_{k_F} \cdot k_F.  \label{Lexp}
\end{eqnarray}
It is worth mentioning that the analytical expressions in
Eq.(\ref{Esymexp}) and Eq.(\ref{Lexp}) are valid at any density
and the values of both $E_{sym}(\rho)$ and $L(\rho)$  can be
easily calculated simultaneously once the single-particle
potential is known. The most critical advantage of the expression
in Eq.(\ref{Lexp}) is that it allows us to determine the $L(\rho)$
directly from the value and momentum dependence of the single
nucleon potential at $\rho$. Essentially, this enables us to
translate the task of determining the density slope of the
symmetry energy into a problem of finding the momentum dependence
of the $U_0(\rho,k)$ and $U_{sym}(\rho,k)$. Very fortunately, the
latter is accessible at least at normal density from
phenomenological optical model analyses of nucleon-nucleus
scatterings, (p, n) charge exchange reactions and the
single-particle energy levels probed by pickup and stripping
reactions \cite{Hodg,Sat79,Mah91}. The fundamental reason for this
interesting feature can be understood mathematically from
inspecting the Eqs.(\ref{chemUn}) and (\ref{chemUp}). It is seen
that the right hand side of these equations relates to the density
slope of the symmetry energy, while the left hand side relates to
the momentum slope of single nucleon potentials when both the
$t(k_F^{n/p})$ and $U_{n/p}$ are expanded as power series of
$\delta$ through the neuron/proton Fermi momentum.

To evaluate the symmetry energy and its density slope at $\rho_0$
using the Eqs.(\ref{Esymexp}) and (\ref{Lexp}), we need to know the
momentum dependence of both the $U_0(\rho_0,k)$ and
$U_{sym}(\rho_0,k)$. This information can be obtained from the
energy dependence of the depth of the real part of the Global
Optical Potential (GOP). The latter is normally parameterized in the
Woods-Saxon form. Over the last few decades, great progress has been
made in developing the unified GOP for both nuclear structure and
reaction studies. At negative energies the GOP can be constrained by
single-particle energies of bound states while at positive energies
it is constrained by nuclear reaction data \cite{Hodg,Sat79,Mah91}.
In the present study, we shall take the full advantage of the
systematics accumulated in this field. The momentum dependence of
$U_0(\rho,k)$ is conventionally taken into account by using the
nucleon effective mass approximation. In terms of the effective
k-mass $m^*$ defined by $ m^*/m=[1+\frac{m}{k_F} \frac{\partial
U_0}{\partial k}|_{k_F} ]^{-1}$, the $E_{sym}(\rho)$ and $L(\rho)$
can be rewritten as
\begin{eqnarray}
 && E_{sym}(\rho) = \frac{1}{3} \frac{\hbar^2 k_F^2}{2 m^*} +
\frac{1}{2} U_{sym}(\rho,k_F) \label{Esymexp2}
\\
&& L(\rho) = \frac{2}{3} \frac{\hbar^2 k_F^2}{2 m^*} + \frac{3}{2}
U_{sym}(\rho,k_F) + \frac{\partial U_{sym}}{\partial k}|_{k_F}
 k_F. \label{Lexp2}
\end{eqnarray}
For the present study, we only need to know the $m^*$ and
information about the $U_{sym}(\rho,k)$ at $\rho_0$ and the
corresponding Fermi momentum ($k_F=1.36 $ fm$^{-1}$). For the
nucleon effective mass, we adopt the value of $m^*/m=0.7\pm0.05$
\cite{jamo} widely used in the literature, see, \textit{e.g.}
\cite{Bert88}. This value is consistent with the
$U_0(\rho_0,E)=-(50.0-0.30E)$ from the global optical model analysis
of nucleon-nucleus scattering experiments \cite{Hodg,Hama} and the
microscopic many-body calculations \cite{frie,van}. Since the $U_0$
from data analyses is usually expressed as a function of nucleon
energy $E$, a dispersion relation determined by
$U_0(\rho_0,k)=-\{50.0-0.30 [t(k)+U_0(\rho_0,k)]\}$ has to be used
to obtain the $m^*/m=0.7$.

\begin{figure}[htb]
\centering
\includegraphics[width=7cm]{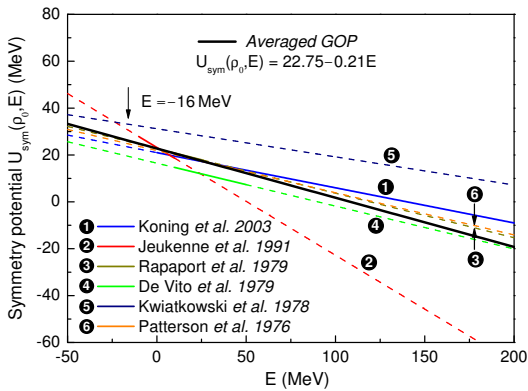}
\caption{(Color online) Energy dependence of the nuclear symmetry
potential $U_{sym}(\rho_0,E)$ at normal density from different
global optical model fits. The solid lines are in the energy
ranges where the original analyses were made while the dashed
parts are extrapolations.}
\end{figure}

The symmetry potential $U_{sym}(\rho_0,E)$ can be deduced from
studying the isospin dependence of the optical potential
\cite{Hodg,Sat69,Sat79,Kho07} using 1) elastic scattering of a
proton and a neutron from the same target at about the same beam
energy after correcting the Coulomb energy for protons; 2) proton
scattering at the same energy on several isotopes of the same
element; 3) (p, n) charge exchange reaction between isobaric
analog states. Since the 1960s, there are many sets of GOPs
deduced from phenomenological model analyses of the world data
existing at the time \cite{gom1,gom2}. While some of the analyses
assumed an energy independent symmetry potential, see,
\textit{e.g.}, \cite{Hodg,gom2}, a significant number of studies
considered the energy dependence \cite{gom1}. We also notice here
that there are also some local/global phenomenological optical
potentials with mass-dependence and semi-microscopic optical
potentials involving complex nuclear structure calculations (see
Ref.\cite{wep} and references therein). Here we only use results
from energy dependent studies. In these analyses the symmetry
potentials are normally described by using a linear function of
the form $U_{sym}(\rho_0,E)=a_{sym}-b_{sym}E$. Shown in Fig.\ 1
are all the energy dependent symmetry potentials that we can find
in the literature \cite{gom1}. While some of them were extracted
from data in the same energy range, somewhat different kinds of
data and models were used. All together, they represent
parameterizations constrained by the world data on nucleon-nucleus
scatterings, (p, n) charge exchange reactions and single-particle
energies of bound states. Assuming that these various global
energy dependent symmetry potentials are equally accurate and all
have the same predicting power beyond the original energy ranges
in which they were analyzed, an averaged symmetry potential of
\begin{equation}
U_{sym}(\rho_0,E)=22.75-0.21E\label{Ubest}
\end{equation}
is obtained (thick solid line in Fig.\ 1). Besides determining the
values of $E_{sym}(\rho_0)$ and $L(\rho_0)$, we expect this
expression to be useful for calibrating $U_{sym}(\rho,E)$ at
abnormal densities predicted by various many-body theories. Under
the assumptions mentioned above, it represents the best fit to the
global symmetry potentials constrained by the world data up to date.
With this best estimate for the energy dependent symmetry potential
at $\rho_0$, it is then straightforward to get the best estimates
for the $E_{sym}(\rho_0)$ and $L(\rho_0)$ using Eq.(\ref{Esymexp2})
and Eq.(\ref{Lexp2}), respectively. Because $E=-$16 MeV at $\rho_0$,
the corresponding symmetry potential is then $U_{sym}(\rho_0,k_F)=
22.75-0.21\times (-16) =26.11$ MeV. With this we obtain the
$E_{sym}(\rho_0)= 31.3$ MeV. This value agrees remarkably well with
that deduced from the experimental binding energies of nuclei.
Similarly, a value of $L(\rho_0)=52.7 \textrm{\,\,MeV}$ is obtained
from Eq.(\ref{Lexp2}).

\begin{figure}[htb]
\centering
\includegraphics[width=7cm]{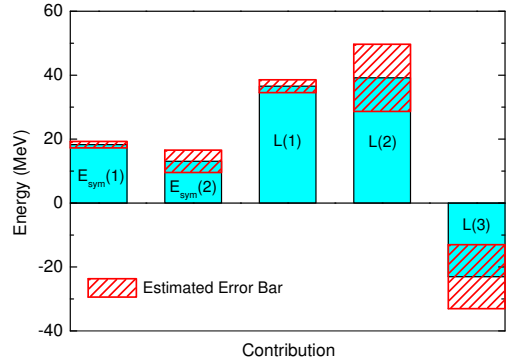}
\caption{(Color online) The magnitude of each term in the nuclear
symmetry energy $E_{sym}(\rho_0)$ and its density slope
$L(\rho_0)$ at normal nuclear density.}
\end{figure}

At the mean-field level, isospin effects are determined by the
quantity $U_{sym}\cdot\delta$. Since the latter is generally much
smaller than $U_0$ \cite{Sat69,Kho05}, it is no surprise that there
are large discrepancies among results from different analyses. There
are thus also probably large error bars associated with our best
estimates for the $E_{sym}(\rho_0)$ and $L(\rho_0)$. However, due to
the wide range of experiments involved and the various methods used
in analyzing the data, it is impossible for us to calculate
accurately these error bars. Nevertheless, it is instructive to
estimate the uncertain ranges of each term in calculating the
$E_{sym}(\rho_0)$ and $L(\rho_0)$. By using the ranges of the
parameters $a_{sym}$ and $b_{sym}$ shown in Fig.1, the uncertainties
of all terms contributing to the $E_{sym}(\rho_0)$ and $L(\rho_0)$
are obtained and shown in Fig. 2. The $E_{sym}(1)=\frac{1}{3}
\frac{\hbar^2 k_F^2}{2 m^*}$ denotes the kinetic energy term with
the effective mass $m^*$ and the $E_{sym}(2)=\frac{1}{2}
U_{sym}(\rho_0, k_F)$ is the symmetry potential contribution. It is
seen from Fig.\ 2 that the two terms are comparable. The uncertainty
for calculating $E_{sym}(1)$ is due to the effective mass $m^*$.
With the $m^*/m=0.70\pm0.05$ we adopted \cite{jamo}, an error bar of
$\pm 1.0$ MeV is obtained for the $E_{sym}(1)$, as marked by the red
box. From the results shown in Fig.1, the symmetry potential
$U_{sym}(\rho_0,k_F)$ at the Fermi momentum $k_F$ is found to lie in
the range of $26.11\pm 7.0$ MeV. So, put all together the error bar
of $E_{sym}(\rho_0)$ is approximately $\pm (1.0+3.5)=\pm4.5$ MeV.
The $L(\rho_0)$ has three terms $L(1)=\frac{2}{3} \frac{\hbar^2
k_F^2}{2 m^*}$, $L(2)=\frac{3}{2} U_{sym}(\rho_0,k_F) $ and $L(3)=
\frac{\partial U_{sym}(\rho,k)}{\partial k}|_{k_F} k_F$. While both
the $L(1)$ and $L(2)$ are positive, the $L(3)$ is negative because
of the decreasing symmetry potential with increasing energy. Similar
to estimating the error bars for the $E_{sym}(1)$ and $E_{sym}(2)$,
an error bar of $\pm2.0$ MeV is estimated for $L(1)$ and $\pm10.5$
MeV for $L(2)$. To evaluate the error bar for $L(3)$, we estimated
an uncertainty of $0.21\pm 0.1$ for $b_{sym}$ from Fig.1. Then, an
error bar of $\pm10$ MeV is obtained for $L(3)$. Thus, the total
error bar of $L(\rho_0)$ is approximately $\pm$ 22.5 MeV.

In summary, using the Hugenholtz-Van Hove theorem, it is shown
that both the symmetry energy and its density slope at normal
density are completely determined by the global nucleon optical
potentials. Using the nucleon effective mass $m^*/m=0.7$ and the
nuclear symmetry potential $U_{sym}(\rho_0,E)=22.75-0.21E$
obtained from averaging all phenomenological isovector nucleon
potentials constrained by world data up to date, the best
estimates of the symmetry energy and its density slope at $\rho_0$
are found to be $E_{sym}(\rho_0)=31.3$ MeV and $L(\rho_0)=52.7$
MeV, respectively. Uncertainties involved in the estimates are
discussed.

This work is supported in part by the US National Science
Foundation grants PHY-0757839, the Research Corporation under
grant No.7123, the Texas Coordinating Board of Higher Education
grant No.003565-0004-2007, the National Natural Science Foundation
of China grants 10735010,10775068,10805026 and 10975097, Shanghai
Rising-Star Program under grant No. 06QA14024, the National Basic
Research Program of China (2007CB815004 and 2010CB833000).

\end{document}